\documentclass[aps,prd,twocolumn,showpacs,floats,floatfix,letterpaper,nofootinbib,superscriptaddress,]{revtex4}

\input epsf
\usepackage{graphicx}
\usepackage{color}
\usepackage{mathtools}
\usepackage{mathbbol}
\usepackage{amssymb}
\usepackage{cancel}

\newcommand{\be}{\begin{equation}}
\newcommand{\ee}{\end{equation}}
\newcommand{\beq}{\begin{equation}}
\newcommand{\eeq}{\end{equation}}
\newcommand{\bea}{\begin{eqnarray}}
\newcommand{\eea}{\end{eqnarray}}

\newcommand{\rme}{\textrm{e}}

\newcommand{\bs}{\boldsymbol}

\newcommand{\lsim}{\mathrel{\hbox{\rlap{\lower.55ex\hbox{$\sim$}} \kern-.3em \raise.4ex \hbox{$<$}}}}
\newcommand{\gsim}{\mathrel{\hbox{\rlap{\lower.55ex\hbox{$\sim$}} \kern-.3em \raise.4ex \hbox{$>$}}}}

\begin{document}

\title{Flaring of tidally compressed dark-matter clumps}

\author{Yacine Ali-Ha\"imoud}
\affiliation{Department of Physics and Astronomy, Johns Hopkins University, Baltimore, MD 21218, USA}
\author{Ely D. Kovetz}
\affiliation{Department of Physics and Astronomy, Johns Hopkins University, Baltimore, MD 21218, USA}

\author{Joseph Silk}
\affiliation{Department of Physics and Astronomy, Johns Hopkins University, Baltimore, MD 21218, USA}
\affiliation{Institut d'Astrophysique de Paris (UMR7095: CNRS \& UPMC- Sorbonne Universities), F-75014, Paris, France}
\affiliation{AIM-Paris-Saclay, CEA/DSM/IRFU, CNRS, Univ. Paris VII, F-91191 Gif-sur-Yvette, France}
\affiliation{BIPAC, Department of Physics, University of Oxford, Keble Road, Oxford
OX1 3RH, UK}

\begin{abstract}
We explore the physics and observational consequences of tidal compression events (TCEs) of dark-matter clumps (DMCs) by supermassive black holes (SMBHs). Our analytic calculations show that a DMC approaching a SMBH much closer than the tidal radius undergoes significant compression along the axis perpendicular to the orbital plane, shortly after pericenter passage. For DMCs composed of self-annihilating dark-matter particles, we find that the boosted DMC density and velocity dispersion lead to a flaring of the annihilation rate, most pronounced for a velocity-dependent annihilation cross section. If the end products of the annihilation are photons, this results in a gamma-ray flare, detectable (and possibly already detected) by the Fermi telescope for a range of model parameters. If the end products of dark-matter annihilation are relativistic electrons and positrons and the local magnetic field is large enough, TCEs of DMCs can lead to flares of synchrotron radiation. Finally, TCEs of DMCs lead to a burst of gravitational waves, in addition to the ones radiated by the orbital motion alone, and with a different frequency spectrum. These transient phenomena provide interesting new avenues to explore the properties of dark matter.

\end{abstract}

\date{\today}

\maketitle

\section{Introduction}

In the quest for the nature of dark matter (DM), cosmologists have set increasingly stringent limits on its properties, shrinking ever further the parameter space in which it is allowed to exist. Yet, despite decades of research, the only characteristics of DM that are robustly measured to date are its mean cosmological abundance and the amplitude of primordial density perturbations on scales larger than a few comoving Mpc. Identifying the DM is a fundamental problem of modern physics; it is therefore important to explore every possible phenomenon that could help characterize it.

Dark matter is expected to be clumpy on small scales \cite{Silk_1993, Berezinsky_14}, even in the simplest scenario where one extrapolates to very small scales the primordial power spectrum measured on large scales by cosmic microwave background (CMB) experiments \cite{Planck_2015_XIII}. Since the amplitude and character of primordial fluctuations on scales smaller than a few Mpc are very poorly constrained, it is also possible that there exists a population of very dense dark-matter clumps (DMCs). These could have formed shortly after matter-radiation equality from adiabatic perturbations with overdensity $\delta \lesssim 0.3$. They could also have formed earlier on as a result of accretion onto primordial black holes \cite{Bertschinger_85}, the collapse of primordial isocurvature perturbations \cite{Kolb_1994}, topological defects or phase transitions \cite{Berezinsky_10}. Analytic self-similar solutions for radial infall predict a cuspy power-law density profile \cite{Fillmore_84, Bertschinger_85}. Of course the collapse is never perfectly radial and the density does not increase to arbitrarily large values in the center, so DMCs are expected to have cores of a small fraction of their virial radii \cite{Bringmann_12}. If they are sufficiently dense and gravitationally bound, these cores can survive destruction due to interactions with stars and the tidal field of the galaxy that harbors them \cite{Berezinsky_06, Bringmann_12}. 

Various predicted signals have been proposed to test the properties of DMCs, such as the gamma ray emission resulting from annihilating DM \cite{Berezinsky_03, Scott_09} or microlensing events \cite{Ricotti_09}. In this paper, we explore for the first time the observational consequences of a phenomenon that is likely to bring the demise of some DMCs: strong tidal interaction with a supermassive black hole (SMBH).

The centers of most galaxies are believed to harbor a SMBH, cf. \cite{Kormendy_13}. It is well known that stars whose trajectories pass close enough to a SMBH can be tidally disrupted, leading to bright transient events that are actively being studied \cite{Arcavi_14}. A perhaps less known phenomenon is the tidal \emph{compression} of stars by SMBHs \cite{Carter_82, Stone_2013}. The physical mechanism is rather simple: while tidal forces stretch the star along the axis joining its center of mass to the SMBH, they compress it along the two perpendicular directions. 

In this paper, we study, for the first time, the consequences of the equivalent phenomenon for DMCs. We compute the net compression of a DMC approaching a SMBH beyond the Roche radius. We find that for orbits penetrating deep inside the Roche radius $R_t$, the DMC is compressed shortly after pericenter passage by a factor of order $\beta = R_t/R_p \gg 1$, where $R_p$ is the distance to the black hole at pericenter. The compression takes place on a timescale of order $\beta^{-2}$ times the DMC dynamical timescale. Tidal compression events (TCEs) of DMCs therefore lead to \emph{flares}, either of gamma rays if the final products of DM annihilation are photons, or of synchrotron radiation if DM particles annihilate via leptonic channels into relativistic electrons and positrons and the local magnetic field is strong enough. Such events also lead to bursts of gravitational waves.  

A prediction of the rate of TCEs of DMCs from first principles is highly model-dependent and would require treating a chain of complex processes. First, one would need to predict or parameterize the abundance, mass function and profiles of primordial DMCs. Second, one would have to compute the survival rates of DMCs through hierarchical structure formation and tidal interactions with the galactic field and passing stars. Finally, one would need to estimate the rate at which their orbits are deflected into the SMBH ``loss cone", i.e. close enough that they can be tidally compressed \cite{Magorrian_99, Wang_04}. This problem may be complicated by the fact that DMCs may be less dense than stars, and more likely to be disrupted by close interactions than to be deflected. In this first exploratory study, we study TCEs from a phenomenological point of view, taking the DMC mass, density and pericenter distance as free parameters, and defer an estimation of the event rate to future work. We hope that our results motivate more detailed follow-up, especially using appropriate numerical simulations.

The structure of this paper is as follows. In Section \ref{sec:theory}, we lay out our model for DMCs and compute their net compression during a TCE. In Section \ref{sec:observation}, we calculate the predicted flux and duration of gamma-ray or synchrotron flares resulting from dark matter annihilation, and compare them with some existing observations. We also estimate the gravitational wave signal.  
In Section \ref{sec:discussion} we discuss possible improvements and generalizations of the model. We conclude in Section \ref{sec:conclusion}.
 
\section{Theory}
\label{sec:theory}

\subsection{Dark matter clump properties}

The density profile of a DMC  depends on the details of its formation process \cite{Kolb_1994, Berezinsky_13}. As shown for example in Ref.~\cite{Kolb_1994}, if the DMC forms from a primordial adiabatic density fluctuation with $\delta \lesssim 0.3$ (larger adiabatic fluctuations would form primordial black holes \cite{Carr_2005}), and then collapses at the earliest at matter-radiation equality, its final density is typically $\sim 100$ times the background density at collapse, $\rho_{\rm DM} \lesssim 10^{-18}$ g/cm$^{3}$, which is rather small.

However, a generic outcome of analytic and numerical calculations is that DMCs have a cuspy power-law profile \cite{Fillmore_84, Bertschinger_85}. This power law does not continue to arbitrarily small radii: at some point the density is expected to flatten out to a constant value, which we denote by $\rho_{\rm cl}$. Existing dark-matter-only\footnote{Of course hydrodynamic simulations including baryons can resolve a core \cite{Schaller_16, Calore_15}, but the DMCs we consider are too small to retain any baryons.} simulations do not have sufficient resolution to capture the transition from a cuspy power-law profile to the core (see e.g.~\cite{Ishiyama_10, Schaller_16}). The core size and central density are therefore highly uncertain \cite{Berezinsky_08}. Several processes can prevent the density from reaching arbitrarily high values at the center \cite{Berezinsky_13}. If the DM particles have some initial random velocities, either thermal or turbulent, the decrease of the coarse-grained phase-space density imposes a lower bound on the core radius \cite{Tremaine_79}. If the dark matter self-annihilates, one can set an upper limit to the core density by requiring that it does not entirely self-annihilate within a Hubble time \cite{Ullio_02}, or, more realistically, within a dynamical time as the core should be refilled on that time-scale \cite{Berezinsky_14}. The latter requirement implies
\beq
\rho_{\rm cl} \frac{\langle \sigma v \rangle}{m_{\chi} } \lesssim T_{\rm cl}^{-1}, \label{eq:rhocl_max}
\eeq
where $\langle \sigma v \rangle$ is the velocity-averaged annihilation cross section times velocity, $m_{\chi}$ is the mass of the DM particle, and $T_{\rm cl}$ is the dynamical timescale of the self-gravitating DMC, 
\be
T_{\rm cl} \equiv \sqrt{\frac{3}{4 \pi G \rho_{\rm cl}}} \approx 0.5 \textrm{ hour} \left(\frac{\rho_{\rm cl}}{1~ \textrm{g cm}^{-3}} \right)^{-1/2}.
\ee 
The maximum clump density for self-annihilating DM is therefore
\bea
\rho_{\rm cl} &\lesssim& \frac{4 \pi G}{3} \left( \frac{m_\chi}{\langle \sigma v \rangle} \right)^{2}\nonumber\\
&\approx& 10 ~\textrm{g cm}^{-3} ~\left( \frac{\langle \sigma v \rangle}{3{\times}10^{-26} \textrm{cm}^3 \textrm{s}^{-1}} \right)^{-2} \left(\frac{m_\chi}{100~\textrm{GeV}}\right)^2,~~~\label{eq:rho_max}
\eea
of order the mean Earth density for a characteristic thermal relic velocity-averaged cross section \cite{Jungman_96} and 100 GeV dark-matter particle. We note, however, that unless the DM annihilates through an $s$-wave interaction ($\sigma v = $ constant), the value of $\langle \sigma v \rangle$ for the considered DMC depends on its internal velocity dispersion and need not be equal to the thermal relic value.

Clearly, a minimum core density is required for the DMC to survive disruption by stellar encounters or galactic tides before reaching the immediate vicinity of the SMBH. The survival rate of DMCs is a subject of active work and debate \cite{Zhao_05, Berezinsky_06, Green_07, Berezinsky_08}, and we shall not venture an estimate of the minimum DMC density here. We shall instead keep $\rho_{\rm cl}$ as a free parameter, bounded from above as in Eq.~\eqref{eq:rho_max} (when considering self-annihilating DM), and defer a detailed study of allowed values for future work.

In what follows we shall only focus on the DMC's \emph{core}. We denote by $M_{\rm cl}$ the DMC's mass (more precisely, its core's mass). The characteristic clump radius $R_{\rm cl}$ is defined as
\beq
R_{\rm cl} \equiv \left(\frac{3 M_{\rm cl}}{4 \pi \rho_{\rm cl}}\right)^{1/3} \approx 6 \times 10^{-5}  (M_{\rm cl, \odot})^{1/3} \rho_{10}^{-1/3} \textrm{pc},
\eeq
where from this point on, we use $M_{\rm cl, \odot} \equiv M_{\rm cl}/M_{\odot}$ and $\rho_{10} \equiv \rho_{\rm cl}/(10^{-10}$ g/cm$^3)$. This particular normalization is arbitrary, but all our expressions remain completely general.

\subsection{Infall into a SMBH}

If the DMC is orbiting a SMBH with mass $M_{\rm BH}$, then the tidal radius, at which the tidal field of the SMBH overcomes the DMC self gravity, is given by
\be
R_t \equiv \left(\frac{3 M_{\rm BH}}{4 \pi \rho_{\rm cl}}\right)^{1/3} \approx 6 \times 10^{-3}  ~M_{\rm BH, 6}^{1/3}~ \rho_{10}^{-1/3} \rm pc,
\ee
where $M_{\rm BH, 6} \equiv M_{\rm BH}/(10^6 M_{\odot})$. This is typically much larger than the Schwarzschild radius of the SMBH, $R_{\rm BH} = 2 G M_{\rm BH}/c^2 \approx 10^{-7} M_{\rm BH, 6}$ pc, as long as 
\beq
\rho_{\rm cl} \ll 2 \times 10^4 ~\textrm{g cm}^{-3}~ M_{\rm BH, 6}^{-2}.
\eeq
We assume that the center of mass of the DMC is on a parabolic orbit around the SMBH, with pericenter $R_p$. Its distance from the SMBH is given by
\be
R = \frac{2R_p}{1+\cos{f}},
\ee
where the angle $f$ is the true anomaly. Following Ref.~\cite{Carter_82}, we define the penetration factor 
\be
\beta \equiv \frac{R_t}{R_p}.
\ee
Note that assuming $R_p \geq R_{\rm BH}$ imposes an upper bound on the penetration factor $\beta \leq \beta_{\max}$, with
\be
\beta_{\max} \approx 6 \times 10^4 ~M_{\rm BH, 6}^{-2/3} ~\rho_{10}^{-1/3}.
\ee

Once the DMC is deep enough inside the tidal radius, tidal forces dominate over its self gravity. The DMC is then effectively gravitationally unbound, and all of its particles are in free fall. The DMC is then tidally stretched along the axis joining its center of mass to the SMBH and tidally compressed along the perpendicular directions. Unlike stars, for which internal pressure eventually halts tidal compression, the DMC can a priori be compressed to arbitrarily large densities. Initial turbulent velocities will however prevent the DMC from reaching infinite densities, as they lead to a de-synchronization of orbit crossings \cite{Stone_2013}.
 
We shall restrict ourselves to penetration factors $\beta \leq (M_{\rm BH}/M_{\rm cl})^{1/3}$ for which the DMC-SMBH separation is always larger than the DMC size, $R_{\rm cl} \ll R$. This allows us to compute the tidal deformation of the clump perturbatively by Taylor-expanding the equations of motion of its constituent mass elements around the trajectory $\bs{R}$ of the center of mass. It would be interesting to generalize our analysis to larger penetration factors, and we defer this to future work.

We define $\bs{r}$ to be the distance of a mass element to the center of mass, divided by $R_{\rm cl}$. We also define $\bs{v} = \dot{\bs{r}}$, where an overdot denotes differentiation with respect to the normalized time $t/T_{\rm cl}$. The linearized equation of motion for the dimensionless separation $\bs{r}$ is then
\be
\ddot{\bs{r}} = \left(\frac{R_t}{R}\right)^3 \left( 3(\hat{R} \cdot \bs{r})\hat{R}- \bs{r} \right). \label{eq:ddot_r}
\ee
This can be transformed into a differential equation in $f$ by using 
\be
\frac1{T_{\rm cl}}\frac{d t}{df} = \sqrt{8} \beta^{-3/2} (1 + \cos f)^{-2}. \label{eq:dtdf}
\ee
By virtue of the linearity of Eq.~\eqref{eq:ddot_r}, we may linearly relate the phase-space coordinate $(\bs{r}, \bs{v})$ at true anomaly $f$ to the initial coordinates $(\bs{r}_0, \bs{v}_0)$ at the crossing of the tidal radius, which occurs at true anomaly $f_0 = - \textrm{arccos}(2/\beta -1)$, assuming the transition from the domination of self-gravity to that of tidal forces is instantaneous. Since the system is deterministic, the transformation is invertible and we shall denote by $(\bs{r}_0, \bs{v}_0)_{(\bs{r}, \bs{v}, f)}$ the unique initial conditions leading to $(\bs{r}, \bs{v})$ at $f$.

Liouville's theorem ensures that the phase-space density $\mathcal{F}(\bs{r}, \bs{v})$ is conserved: $\mathcal{F}(\bs{r}, \bs{v}, f) = \mathcal{F}((\bs{r}_0, \bs{v}_0)_{(\bs{r}, \bs{v}, f)}, f_0)$. This allows us to compute the (normalized) density field $\tilde{\rho}(\bs{r}, f)$:
\bea
\tilde{\rho}(\bs{r}, f) &=& \int d^3 v~ \mathcal{F}(\bs{r}, \bs{v}, f) \nonumber\\
&=& \int d^3 v ~\mathcal{F}((\bs{r}_0, \bs{v}_0)_{(\bs{r}, \bs{v}, f)}, f_0). \label{eq:rho}
\eea
To simplify calculations we assume that the initial phase-space density is Gaussian and spherically symmetric, both in position and velocity, and takes the form:
\be
\mathcal{F}(\bs{r}_0, \bs{v}_0, f_0) = \rho_*~ \rme^{- 3 r_0^2/2} \left(\frac{3}{2 \pi}\right)^{3/2} \rme^{- 3 v_0^2/2},\label{eq:F_0}
\ee
with $\rho_{*} \equiv \sqrt{6/\pi} ~R_{\rm cl}^3 ~\rho_{\rm cl}$. This form is in turn factorizable in products of phase-space densities along each axis. Note that the prefactor $\sqrt{6/\pi} \approx 1.38$ is chosen so that the simple relation $M_{\rm cl} = (4 \pi/3) \rho_{\rm cl} R_{\rm cl}^3$ holds. 

We see from Eq.~\eqref{eq:ddot_r} that the evolution of $\bs{r}$ in the orbital plane (component $\bs{r}_{||}$) is decoupled from that in the direction perpendicular to it ($z$-component). The integral in Eq.~\eqref{eq:rho} is therefore factorizable into  in-plane and out-of-plane contributions, which we study separately below.

\subsection{Compression perpendicular to the orbital plane}\label{sec:z}

\begin{figure}[h]
\includegraphics[width=\columnwidth]{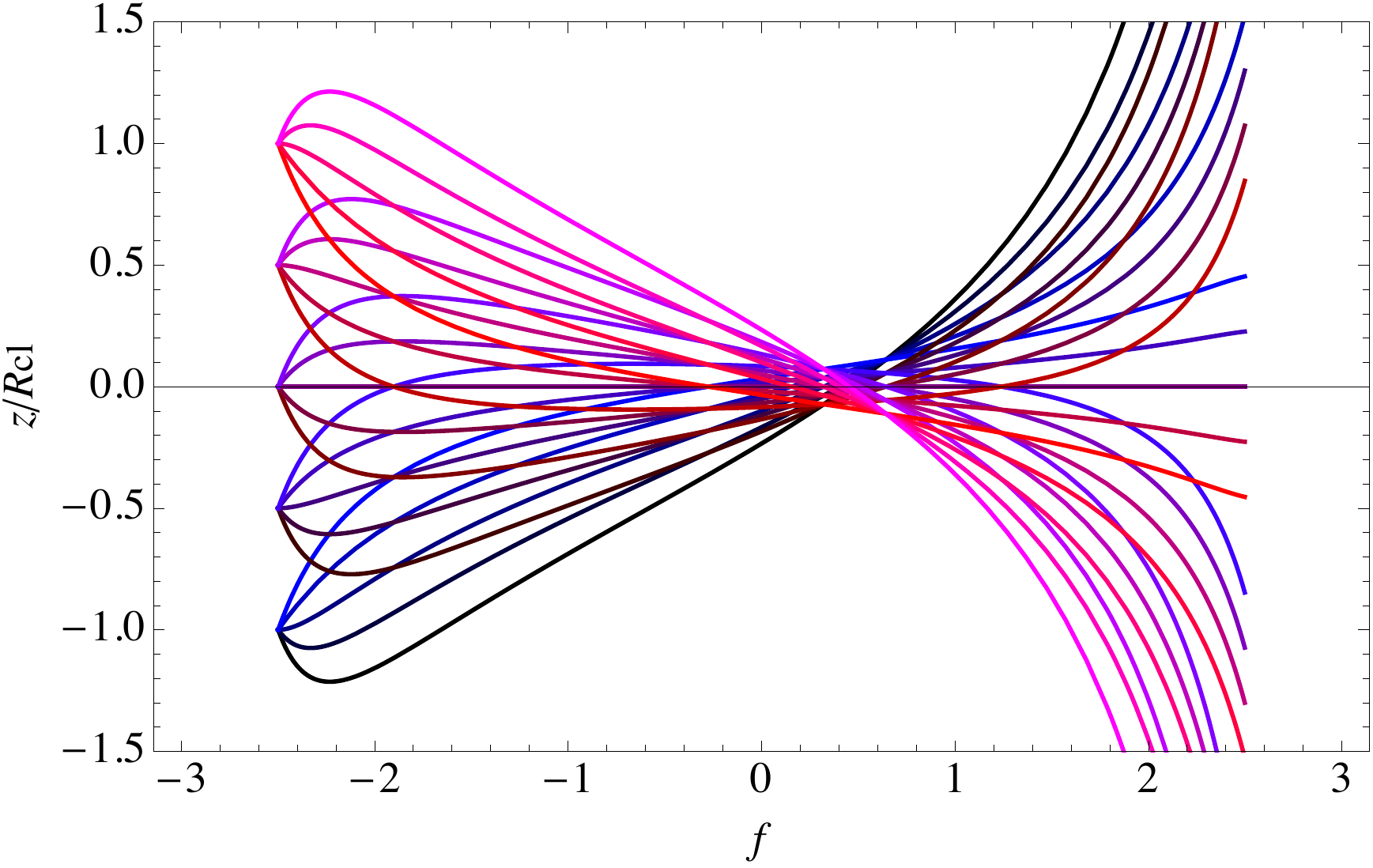}
\caption{Trajectories of test particles with respect to the center of mass of the DMC, perpendicular to the orbital plane, for a penetration factor $\beta = 10$. Specifically, we show trajectories with initial height $z/R_{\rm cl} \in [-1, 1]$ and initial vertical velocity from -1 to 1 times the virial velocity. The curves are started and ended at entry and exit from the tidal radius.}
\label{fig:z_traj}
\end{figure}

\begin{figure}[h]
\includegraphics[width=\columnwidth]{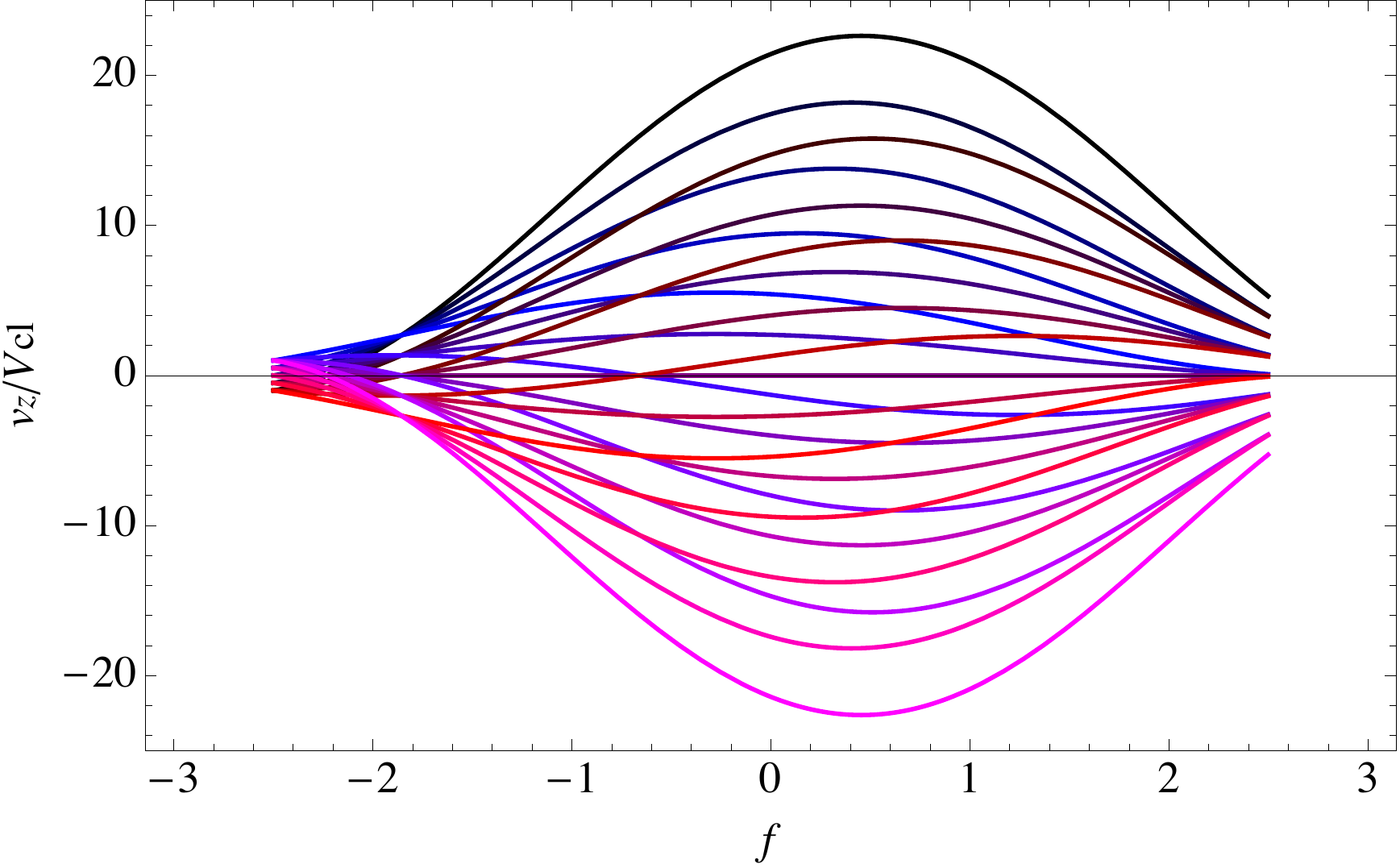}
\caption{Velocities of test particles with respect to the center of mass of the DMC, perpendicular to the orbital plane, for a penetration factor $\beta = 10$. For each curve, the initial conditions match those of the line with identical color in Fig.~\ref{fig:z_traj}}
\label{fig:vz_traj}
\end{figure}

The evolution perpendicular to the orbital plane takes the form
\bea
z(f) &=& a(f) z_0 + b(f) v_{z0}, \label{eq:zofz0}\\
v_z(f) &=& \dot{a}(f) z_0 + \dot{b}(f) v_{z0},
\eea
where $a(f)$ and $b(f)$ are given explicitly in Ref.~\citep{Stone_2013} as functions of the penetration fator $\beta$. We show some example trajectories in Figs.~\ref{fig:z_traj} and \ref{fig:vz_traj}.

\begin{figure}[h]
\includegraphics[width=\columnwidth]{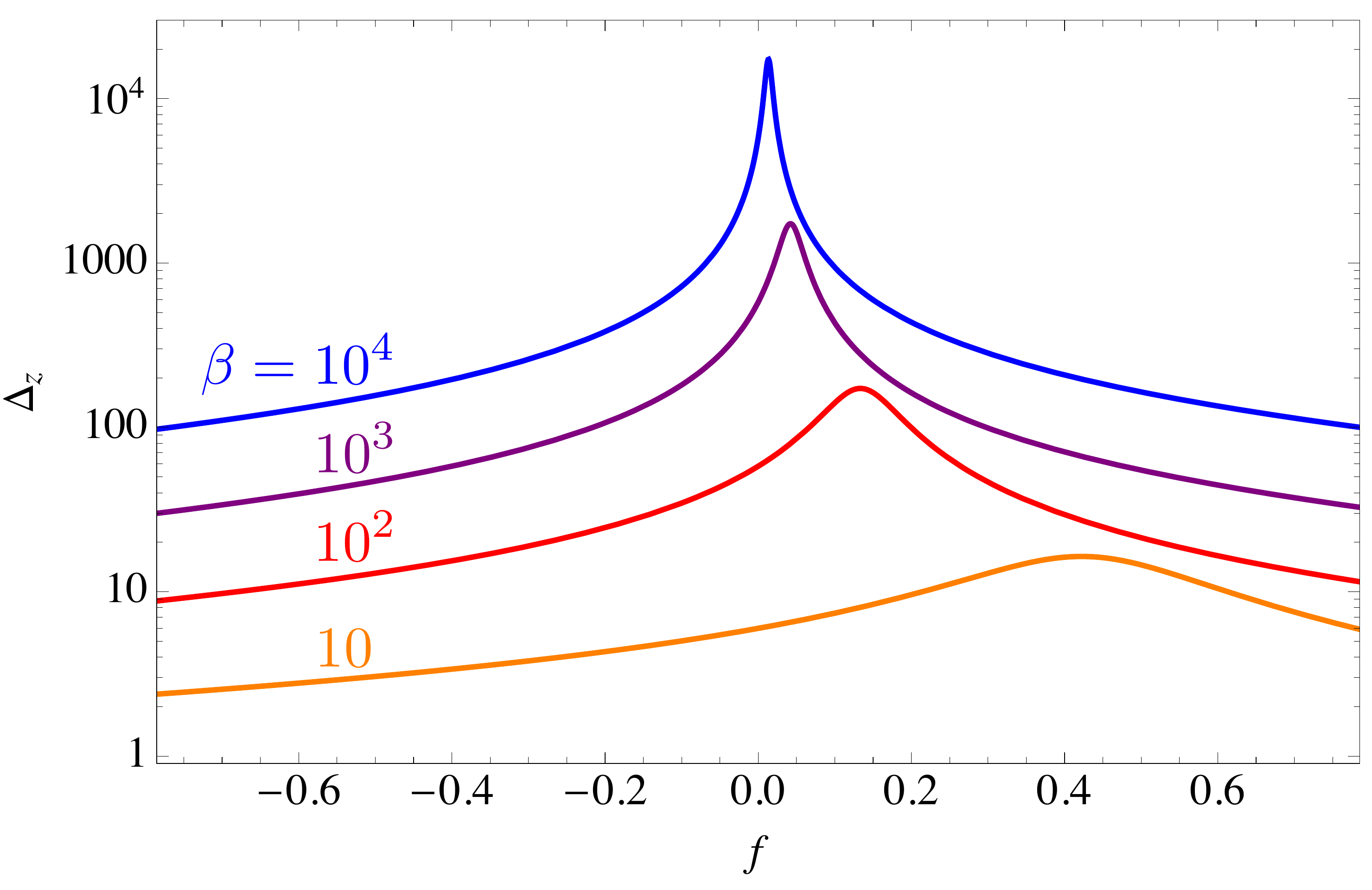}
\caption{Compression factor perpendicular to the orbital plane, as a function of the true anomaly $f$ (focusing on the interval $- \pi/4 < f < \pi/4$), for several values of the penetration factor $\beta$. Pericenter passage occurs at $f = 0$.}
\label{fig:comp_z}
\end{figure}

Since the out-of-plane evolution is separately Hamiltonian, it also satisfies Liouville's theorem, and conserves the phase-space volume. The transformation $(z_0, v_{z0}) \rightarrow (z, v_z)$ therefore has unit determinant ($a \dot{b} - \dot{a} b = 1$) and its inverse is readily computed:
\bea 
z_0 &=& \dot{b}(f) z(f) - b(f) v_z(f), \\
v_{z0} &=& - \dot{a}(f) z(f) + a(f) v_z(f).
\label{eq:z0vz0}
\eea
Assuming $a \neq 0$ we rewrite Eq.~\eqref{eq:zofz0} as 
\be
z_0 = \frac{z}{a} - \frac{b}{a} v_{z0}. 
\ee
The contribution of the $z$-axis to Eq.~\eqref{eq:rho} is
\bea
&&\tilde{\rho}_z(z, f) = \rho_{*} ^{1/3}\sqrt{\frac{3}{2 \pi}} \int d v_z~ \rme^{-3z_0^2/2} \rme^{- 3 v_{z0}^2/2} \nonumber\\
&&= \rho_{*} ^{1/3}\sqrt{\frac{3}{2 \pi}} \int \frac{d v_{z0}}{|a|}~ \rme^{-3(z/a - (b/a) v_{z0})^2/2} \rme^{- 3 v_{z0}^2/2}, ~~~
\eea
where in the second line we changed integration variables from $v_z$ to $v_{z0}$ at fixed $z$, hence the factor $1/|a|$, obtained from Eq.~(\ref{eq:z0vz0}). Performing the Gaussian integral we arrive at
\be
\tilde{\rho}_z(z, f) = \frac{\rho_{*}^{1/3}}{\sqrt{a^2 + b^2}} \exp\left[{-\frac32 \frac{z^2}{a^2 + b^2}}\right].
\ee
We see that the DMC's density profile along the $z$-axis remains Gaussian (if it was initially so) with a characteristic extent $\sqrt{a^2(f) + b^2(f)}$. We show the characteristic compression factor $\Delta_z \equiv 1/\sqrt{a^2(f) + b^2(f)}$ in Fig.~\ref{fig:comp_z}.

Ref.~\cite{Stone_2013} gives explicit expressions for $a(f)$ and $b(f)$. In the large-$\beta$ limit, they are approximately
\bea
a(f) &\approx& \frac{2 \beta^{-1}}{1 + \cos f}(\cos f - \sqrt{\beta} \sin f),\\
b(f) &\approx& \frac{2 \beta^{-1}}{1 + \cos f}(\sqrt{2} \cos f - \sqrt{\beta/2} \sin f). 
\eea
By Taylor-expanding the compression factor $\Delta_z = 1/\sqrt{a^2 + b^2}$ for $f \ll 1$, we arrive at
\be
\Delta_z(f, \beta) \approx \frac{\beta}{\sqrt{\frac32 \beta f^2 - 4 \sqrt{\beta} f + 3}}, \label{eq:comp_approx}
\ee
which peaks at true anomaly $f_{\max} = 4/(3 \sqrt{\beta})$ with maximum value 
\be
\max[\Delta_z] = \sqrt{3}~\beta,
\ee 
and width (measured between the two passages at half-maximum) 
\beq
\Delta f = \sqrt{8/(3 \beta)}. \label{eq:Delta_f}
\eeq
In order to convert this to a duration, we use the differential relation between time and the true anomaly, Eq.~\eqref{eq:dtdf}, and obtain the following time interval between the two passages at half maximum:
\bea
\Delta t_{\rm tce} \approx \frac{2}{\sqrt{3}}~ T_{\rm cl} \beta^{-2} \approx 6  \textrm{ hours} \ \rho_{10}^{-1/2} (10^2/\beta)^{2}.
\label{eq:deltat}
\eea

\subsection{Deformation in the orbital plane}

The same steps can be followed for the DMC deformation in the orbital plane. Starting from the relation 
\be
\bs{r}_{||}(f) = \bs{A}(f) \bs{r}_{||0} + \bs{B}(f)  \bs{v}_{||0},
\ee
where $\bs{A}(f)$ and $\bs{B}(f)$ are two-by-two matrices, we show in Appendix \ref{app:in-plane} that the contribution of the in-plane axes to Eq.~\eqref{eq:rho} is
\bea
\tilde{\rho}_{||}(\bs{r}_{||}, f) &=&  \frac{\rho_{*}^{2/3}}{\sqrt{\det(\bs{A} \bs{A}^{\rm T} + \bs{B} \bs{B}^{\rm T})}}\nonumber\\
&\times& \exp\left[-\frac32 \bs{r}_{||}^{\rm T} (\bs{A} \bs{A}^{\rm T} + \bs{B}\bs{B}^{\rm T} )^{-1}\bs{r}_{||} \right],\label{eq:rho_inplane}
\eea
where the superscript ``T" denotes the transpose. We see that the isodensity contours are deformed into ellipses. The characteristic elongations are the square roots of the eigenvalues of $\bs{A} \bs{A}^{\rm T} + \bs{B}\bs{B}^{\rm T} $. 

Ref.~\cite{Stone_2013} explicitly provides the components of $\bs{A}(f)$ as a function of $\beta$, but not those of $\bs{B}(f)$. We compute $\bs{B}(f)$ by numerically solving the ODE satisfied by $\bs{r}_{||}(f)$. As a sanity check we have verified that our numerical solution for $\bs{A}$ reproduces that of Ref.~\cite{Stone_2013} and that the determinant of the bloc matrix with rows $(\bs{A}, \bs{B})$ and $(\bs{\dot{A}}, \bs{\dot{B}})$ is unity, as it should.

\begin{figure}[h]
\includegraphics[width=\columnwidth]{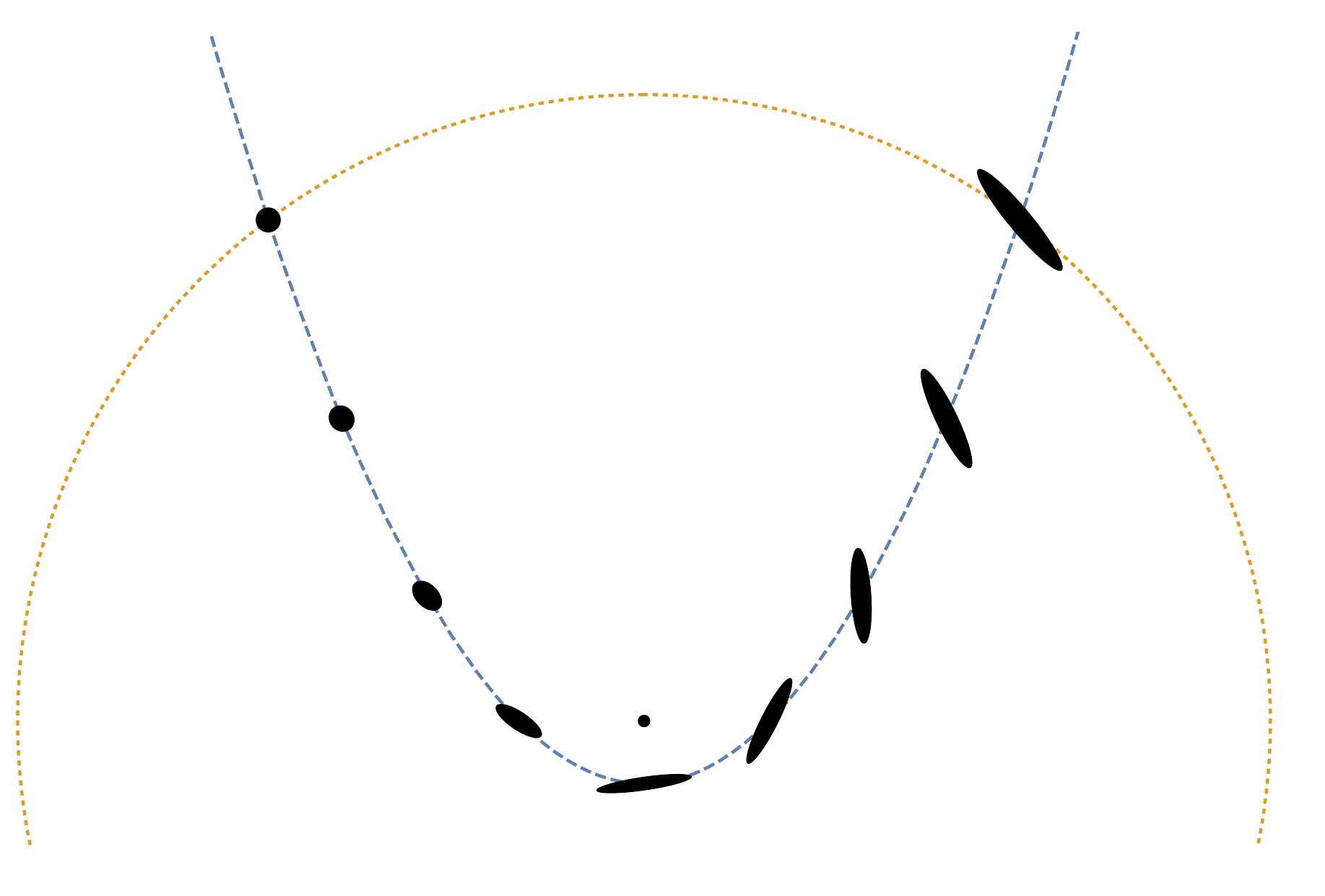}
\caption{Orbital plane deformation of the DMC, with a penetration factor of $\beta=10$. At any given moment, tidal forces tend to stretch the DMC along the axis connecting it to the SMBH and compress it in the perpendicular direction. This figure illustrates the delayed response of the DMC as it moves along its parabolic orbit.}
\label{fig:Trajectory}
\end{figure}

In Fig.~(\ref{fig:Trajectory}), we plot the trajectory of the DMC as it moves around the SMBH, numerically solving for its deformation. We take $\beta= 10$.

\begin{figure}[h]
\includegraphics[width=\columnwidth]{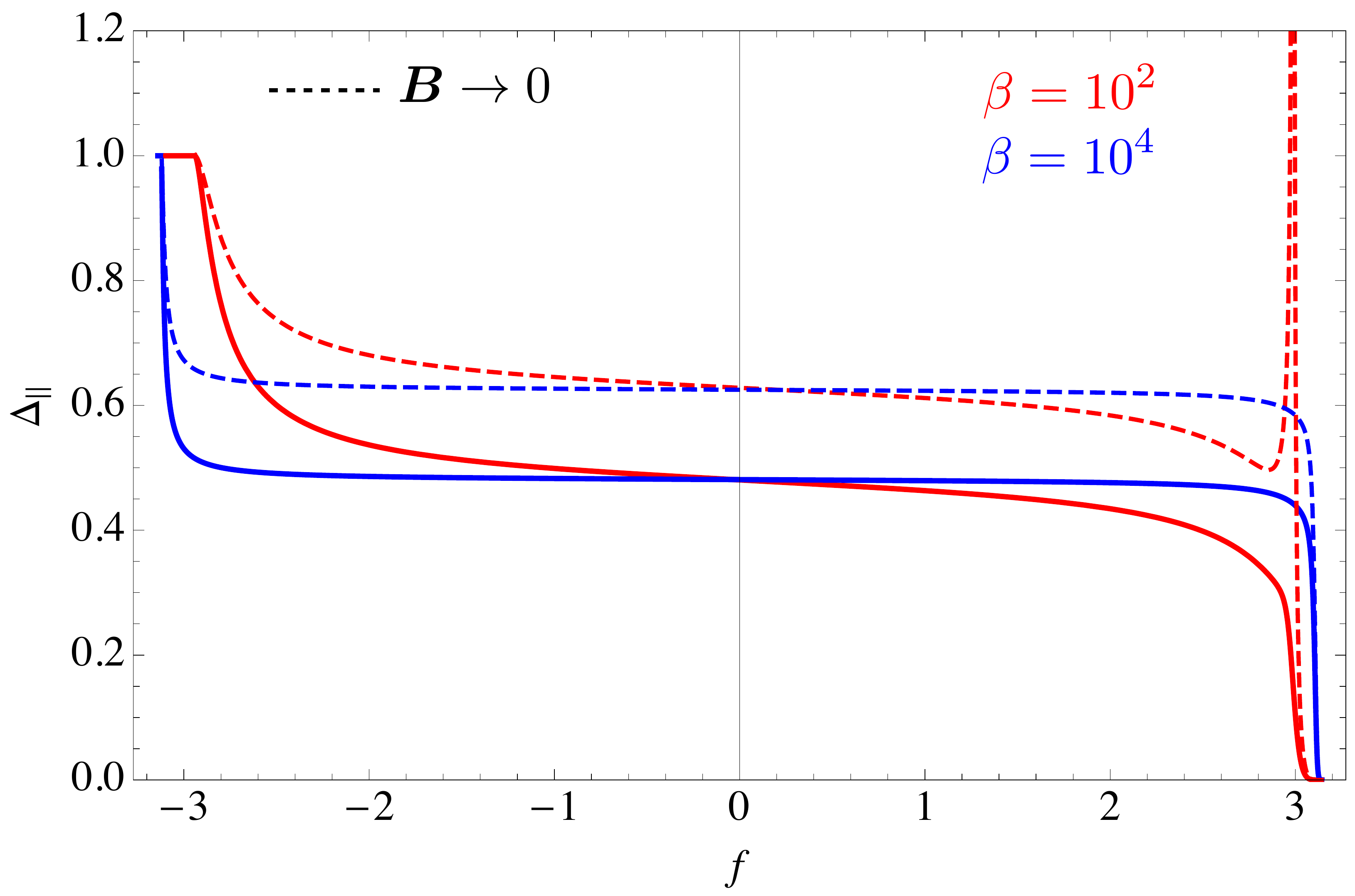}
\caption{Compression factor in the orbital plane, as a function of the true anomaly $f$, for several $\beta= 10^2$ (red) and $10^4$ (blue). The dashed lines illustrate the compression factor that one would obtain if neglecting random velocities (or setting $\bs{B}$ to zero).}
\label{fig:comp_XY}
\end{figure}

We show the in-plane compression factor $\Delta_{||} \equiv 1/\sqrt{\det(\bs{A}\bs{A}^{\rm T}  + \bs{B} \bs{B}^{\rm T} )}$ in Fig.~\ref{fig:comp_XY}. We see that the net result of tidal forces is to stretch the DMC ($\Delta_{||} < 1$). However, this stretching remains of order unity even for large penetration factors, in qualitative agreement with the results of Ref.~\citep{Stone_2013}. We find that accounting for random in-plane motions further reduces the net compression factor (i.e. increase the net stretching) by $\sim 25\%$. The characteristic value of $\Delta_{||}$ near pericenter and for large $\beta$ is $\Delta_{||} \approx 0.5$.

\section{Observable signatures}
\label{sec:observation}

\subsection{Gamma-ray flares from dark-matter annihilation}

\subsubsection{Annihilation rate}

Self-annihilating WIMPs are among the most motivated candidates for dark matter \cite{Jungman_96}. The observable signals resulting from their continuous annihilation near the Galactic center have been studied by several groups \cite{Bertone_02, Regis:2008ij}. The signal can be enhanced if the dark matter is clumpy \cite{Berezinsky_03}. Here we consider a qualitatively new aspect: the transient \emph{flaring} of the annihilation rate that results from the tidal compression of DMCs.

We assume that to lowest order in the velocity, the annihilation cross section has the form $\sigma v = \alpha v^{2\ell}$, where $\alpha$ is a constant. The indices $\ell = 0$  and $\ell = 1$ correspond to $s$-wave and $p$-wave annihilation, respectively \cite{Jungman_96}. The exponent $\ell = 2$ corresponds to a $d$-wave annihilation \cite{Toma_13}. 

We denote the mass of the dark-matter particle by $m_{\chi}$. For $s$-wave annihilation the ratio $p_{\rm ann} \equiv \langle \sigma v \rangle/m_{\chi}$ is constrained by CMB anisotropy measurements to be less than $p_{\rm ann}^{\max} \equiv 3.5 \times 10^{-28}$ cm$^3$ s$^{-1}$ GeV$^{-1}$ (up to an efficiency factor of order unity) \cite{Planck_2015_XIII}.

The rate of annihilation events is
\be
\dot{N}_{\rm ann} = \int d^3 r \frac{\rho^2(\bs{r})}{m_\chi^2} \langle \sigma v_{\rm rel} \rangle, \label{eq:Nann_general}
\ee
where $\langle \sigma v_{\rm rel} \rangle$ is averaged over the relative velocity of two DM particles. At any given position, the latter has a Gaussian distribution with zero mean and covariance matrix whose elements are (in units of the DMC virial velocity dispersion $V_{\rm cl}^2 \equiv G M_{\rm cl}/R_{\rm cl}$):
\bea
\langle v_{\rm rel, z}^2 \rangle &=& \frac23 \frac{1}{b^2(f) + a^2(f)} = \frac23 \Delta_z^2,\\
\textrm{Cov}(\bs{v}_{\rm rel, ||} )&=& \frac23 \left[\bs{\tilde{B}}^{\rm T}\bs{\tilde{B}} +  \bs{\tilde{D}}^{\rm T}\bs{\tilde{D}}\right]^{-1},
\eea
where the matrices $\bs{\tilde{B}}$ and $\bs{\tilde{D}}$ are defined in Eqs~\eqref{eq:tildeB} and \eqref{eq:tildeD}, and the in-plane and out-of-plane relative velocities are uncorrelated. We rewrite
\beq
\langle \sigma v_{\rm rel} \rangle = \langle \sigma v \rangle_0 \times \Delta_v,
\eeq
where $\langle \sigma v \rangle_0$ is the value of $\langle \sigma v_{\rm rel} \rangle$ for the isolated DMC, prior to its entry inside the tidal radius, and $\Delta_v$ is a boost factor due to the velocity dependence of the annihilation rate. Integrating Eq.~\eqref{eq:Nann_general}, the annihilation rate simplifies to
\be
\dot{N}_{\rm ann} = \sqrt{\frac3{4 \pi}}\frac{\langle \sigma v\rangle_0}{m_\chi^2}  \rho_{\rm cl} M_{\rm cl} \times \Delta_z \Delta_{||} \Delta_v. \label{eq:Nann}
\ee
The factor $\Delta_z \Delta_{||} \Delta_v$ is the net boost of the annihilation rate due to tidal compression of the DMC. 

We find that near maximum compression the in-plane relative velocity dispersion [the trace of $\textrm{Cov}(\bs{v}_{\rm rel, ||} )$] is always much smaller than $\langle v_z^2 \rangle$ for large penetration factors. Near maximal compression, we therefore get
\beq
\Delta_v\approx 
\begin{cases} 
1 & s\textrm{-wave}\\
\frac13 \Delta_z^2 & p\textrm{-wave}\\
\frac15  \Delta_z^4 & d\textrm{-wave}
\end{cases}.
\eeq
Therefore, near maximum compression the annihilation rate is enhanced by a factor $\sim \beta^{2\ell + 1}$ with respect to that of a quiescent (non-compressed) DMC. 

The total number of annihilation events during the flare is 
\bea
N_{\rm ann}^{\rm flare} &\approx& \dot{N}_{\rm ann}^{\rm flare} \Delta t_{\rm tce} \nonumber\\
&\approx& 0.5 \frac{\langle \sigma v\rangle_0}{m_\chi^2}  \rho_{\rm cl} M_{\rm cl} \frac{T_{\rm cl}}{\beta} \Delta_v,
\eea
where we used $\Delta_{||} \approx 0.5$ and $\Delta_z \approx \sqrt{3} \beta$ at maximum compression, and $\Delta t_{\rm tce}$ is given by Eq.~\eqref{eq:deltat}.
Numerically, we get, for $\beta \gg 1$, and $\ell = 0, 1$ or 2 (up to a factor of order unity),
\bea
N_{\rm ann}^{\rm flare} \approx 10^{51} \langle \sigma v \rangle_{28} \left(\frac{\textrm{GeV}}{m_{\chi}}\right)^2 \rho_{10}^{1/2} M_{\rm cl, \odot} \times \beta^{2\ell - 1}, \label{eq:Nann_flare}
\eea
where 
\beq
\langle \sigma v \rangle_{28} \equiv \frac{\langle \sigma v \rangle_0}{10^{-28} ~\textrm{cm}^3~ \textrm{s}^{-1}}.
\eeq
We note that for $\beta = 1$ Eq.~\eqref{eq:Nann_flare} gives the number of annihilation events during a DMC dynamical time. 

As an aside, we note that the total time elapsed between entry in the tidal radius and passage at pericenter can be simply obtained by integrating Eq.~\eqref{eq:dtdf} from $f = f_0 = - \textrm{arccos}(2/\beta - 1)$ to 0. In the limit $\beta \gg 1$, we find that this time is $t_{\rm tot} \approx (\sqrt{2}/3) T_{\rm cl}$, of order the DMC dynamical timescale, independently of $\beta$. Therefore as long as the DMC density is below the maximum value given by Eq.~\eqref{eq:rho_max}, it does not entirely annihilate before reaching the pericenter. However, for $p$- and especially $d$-wave annihilation, very dense DMCs could completely annihilate in the flaring event even if their density is below that given in Eq.~\eqref{eq:rho_max}. In this case the light curves that we predict below would be truncated at the time of full DMC annihilation.

\subsubsection{Gamma-ray flux}

The differential gamma-ray flux per energy interval (in photons/s/cm$^2$/GeV) from annihilating dark matter inside an unresolved DMC at distance $d = 100 ~d_{100}$ Mpc from the observer is    
\bea
\frac{d \dot{\phi}}{d E} &=& \frac{d N_{\gamma}}{d E} \frac{\dot{N}_{\rm ann}}{4 \pi d^2}\\
&\approx& 3 \times 10^{-8} ~ \textrm{m}^{-2} \textrm{s}^{-1} \frac{d N_{\gamma}}{d E} \langle \sigma v \rangle_{28}\left(\frac{\textrm{GeV}}{m_{\chi}}\right)^2 \nonumber\\
&&\times    \rho_{10} M_{\rm cl, \odot} (d_{100})^{-2} ~ \Delta_z \Delta_{||} \Delta_v, \label{eq:phi0}
\eea
where $d N_{\gamma}/dE$ is the mean spectrum of gamma-ray photons per annihilation event. We show a few example light curves in Fig.~\ref{fig:lightcurve}.

The total number of photons received per unit area during the flaring event has an energy distribution
\bea
\left(\frac{d \phi}{d E}\right)^{\rm flare}  &=& \frac{d N_{\gamma}}{d E} \frac{N_{\rm ann}^{\rm flare}}{4 \pi d^2}\nonumber\\
&\approx & 10 ~\textrm{m}^{-2} ~ \frac{d N_{\gamma}}{d E} \frac{N_{\rm ann}^{\rm flare}}{10^{51}} (d_{100})^{-2}.
\eea
From Eq.~\eqref{eq:Nann_flare} we see that at equal values of $\langle \sigma v \rangle_0$ for quiescent clumps, $p$-wave and $d$-wave annihilations produce larger flares than $s$-wave annihilations. They are therefore more likely to be observable even if the background gamma ray flux of the possibly numerous quiescent DMCs orbiting the SMBH is undetected.

\subsubsection{TCEs of DMCs as the origin of Fermi flares?}

The Fermi All-Sky Variability Analysis (FAVA) detected 215 flaring gamma-ray sources with photon energies $E \geq 100$ MeV \cite{Fermi_2013}. These sources were selected if their photon count over one week exceeded the expected number of events (rescaled from the total number of events observed over 47 months) by more than 5.5 standard deviations. While known Fermi-LAT sources were found to be associated with the majority of these flaring events, a few tens do not have any known counterpart. Moreover, the associations are purely based on positional coincidence within a broad radius of $\sim$ 1 degree. There is therefore currently no certain explanation for the origin of at least some of these flares, and here we examine whether they could result from TCEs of DMCs.

At high latitudes the threshold for flare identification corresponds to $\sim 100$ photons per week at energies $E \geq 100$ MeV. We assume Fermi's effective area is approximately 0.5 $m^2$. Assuming a $p$-wave annihilation with $\langle \sigma v \rangle_{28} \approx 1$, we see that the tidal compression of a DMC at a distance of 100 Mpc, with mass $M_{\rm cl} \approx 0.1 M_{\odot}$, density\footnote{We note that a power-law primordial power spectrum with adiabatic initial conditions and an amplitude extrapolated from CMB anisotropy measurements cannot lead to DMCs as dense as $10^{-10}$ g/cm$^3$. However, the primordial power spectrum at the relevant very small scales is unconstrained by current observations, and need not have the aforementioned properties. Other scenarios, such as an adiabatic primordial spectrum with localized features or isocurvature fluctuations \cite{Kolb_1994}, could potentially lead to such large overdensities \cite{Berezinsky_13}.} $\rho_{\rm cl} \approx 10^{-10}$ g/cm$^3$ with penetration factor $\beta \approx 100$ could have been at the origin of detected flares. With these parameters the timescale of the flare is several hours, and if the DM particle has mass $m_{\chi} =  1$  GeV and produces $N_{\gamma} = \int dE (d N_{\gamma}/dE) \sim 10$ photons of energy $E \geq 100$ MeV per annihilation, the total number of photons received during the flaring event is of order $\sim 1000$. One may worry that for a TCE to occur a large number of quiescent DMCs orbiting the SMBH is required, and that their combined quiescent emission could overshadow the flare. With the parameters chosen above, we find from Eq.~\eqref{eq:phi0} that $\sim 10^4-10^5$ quiescent DMCs are required to produce $\sim 10^3$ photons during a week. 
These results depend strongly on the assumed microphysical properties of dark matter. Considering d-wave annihilation instead of p-wave, for instance, would result in an enhancement factor larger by $\beta^2$ for the resulting gamma-ray flare from a TCE of DMC, in which case it would take $\sim 10^8-10^9$ quiescent clumps with the same mass and density to produce a similar signal.

\begin{figure}
\centering
\includegraphics[width=0.99\columnwidth]{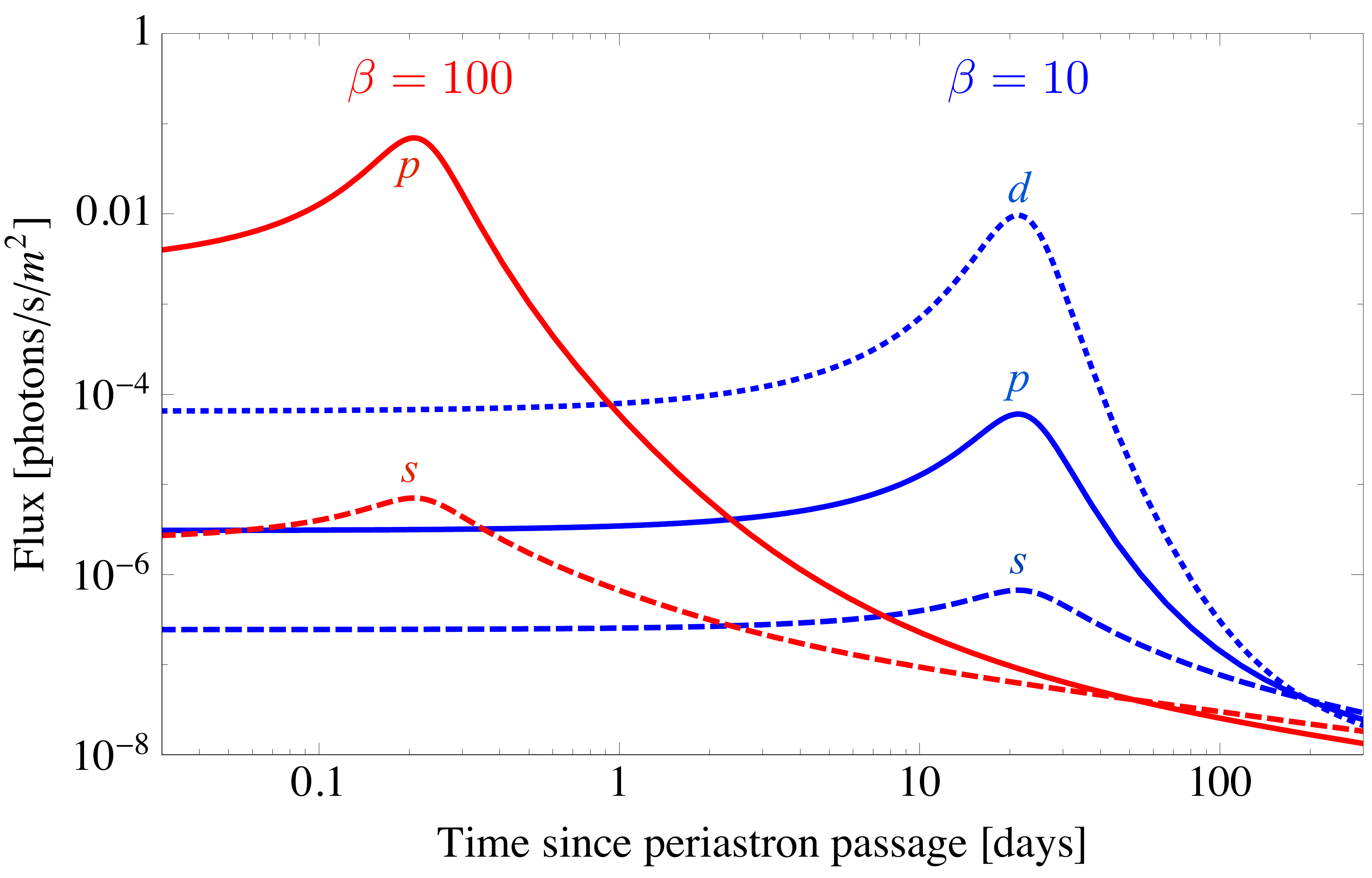}
\caption{Gamma-ray light curves from DM annihilation in tidally compressed DM clumps at a distance of 100 Mpc, assuming $M_{\rm cl} = 0.1 M_{\odot}$, $\rho_{\rm cl} = 10^{-10}$ g cm$^{-3}$, $m_{\chi} = 1$ GeV, and a total number of 10 gamma-ray photons produced per annihilation event. The assumed annihilation cross section is normalized such that $\langle \sigma v \rangle_0 = 3 \times 10^{-28}$ cm$^3$s$^{-1}$ before the DMC enters the tidal radius. The three rightmost (blue) curves correspond to a penetration factor $\beta = 10$, and the two leftmost (red) curves to $\beta = 100$. The labels $s, p, d$ indicate the type of annihilation ($s$-wave, $p$-wave or $d$-wave, respectively).}
\label{fig:lightcurve}
\end{figure}

Given the large freedom allowed by the multiple parameters of the model, it is not surprising that TCEs of DMCs are able to accommodate flares of virtually arbitrary duration and amplitude. However, if this model is to explain some of the Fermi flares (or any other unexplained gamma-ray flaring event), it also makes several additional predictions that would be interesting to check. First and foremost, the energy spectrum of the flares should be universal, as they all arise from the same DM annihilation process. Secondly, we expect flares on a variety of timescales, the statistics of which depends on the distribution of $\beta$, which in turn is set by the loss cone physics \cite{Stone:2014wxa}. If the DMCs have a rather narrow distribution in mass and densities, then we expect a well-defined mapping between the flare amplitude and duration, depending on the type of annihilation: $\dot{\phi}_{\max} \propto (\Delta t)^{-1/2 - \ell}$. Thirdly, the flares should be isotropically distributed, as expected if they take place at cosmological distances. Finally, the light curves of the flares should resemble Fig.~\ref{fig:lightcurve}, though of course a more detailed computation ought to be carried out for a more accurate prediction.

\subsection{Synchrotron flares}

If the end products of dark-matter annihilation are electrons and positrons, they may produce synchrotron radiation due to the magnetic field near the galactic center \cite{Bertone_02, Regis:2008ij}. The synchrotron radiation may not necessarily flare on the timescale of the TCE. Indeed, the energy-loss timescale for relativistic electrons is \cite{Rybicki_book}
\bea
t_{\rm syn} &=& \frac{E}{\dot{E}_{\rm syn}} = \left( \frac43 c \sigma_{\rm T} \frac{B^2}{8 \pi} \frac{E}{(m_e c^2)^2}\right)^{-1}\nonumber\\
&\approx& 1 \textrm{ year} \left(\frac{B}{1 ~\rm G}\right)^{-2} \frac{10~\rm MeV}{E},
\eea
where $\sigma_{\rm T}$ is the Thomson cross section and $B$ is the magnetic field strength. If $t_{\rm syn} < \Delta t_{\rm tce}$, a synchrotron flare would result from the annihilation flare, on a timescale $\Delta t_{\rm tce}$. In the opposite case, the burst of synchrotron radiation would be spread over the timescale $t_{\rm syn}$. 

The amplitude of the magnetic field a fraction of a parsec away from SMBHs is poorly known. Following Ref.~\cite{Aloisio_04}, one may derive an equipartition value for the magnetic field, obtained if the magnetic energy density equals the kinetic pressure. Assuming a Bondi accretion rate of $10^{22}$ g/s $ = 1.6 \times 10^{-4} M_{\odot}$ /year, Ref.~\cite{Aloisio_04} arrive at
\beq
B(R) \approx 0.6 ~M_{\rm BH, 6}^{1/4} \left(\frac{R}{10^{-3} ~\rm pc}\right)^{-5/4} \rm G,
\eeq
where $R$ is the separation from the SMBH. Recent observations of a distant active galactic nucleus (AGN) suggest that the magnetic field could be tens of Gauss, if not significantly larger, 0.01 pc away from the central black hole \cite{Marti-Vidal_15}. While AGNs may have significantly larger magnetic fields than quiescent SMBHs (which would be more appropriate to look for signatures of TCEs of DMCs), this suggests that it is possible that the magnetic field is strong enough that the synchrotron emission could flare on a short timescale. 

To obtain a simple estimate of the synchrotron flux, let us assume for definiteness that $t_{\rm syn} >\Delta t_{\rm tce}$. We denote by $dN_e/dE$ the average electron--positron spectrum per annihilation. Shortly after the TCE, and before the electron energies are significantly dissipated by synchrotron radiation (i.e. if $t < t_{\rm syn}$ for the energies considered), the total synchrotron flux received at a distance $d$ per frequency interval is 
\beq
\frac{d F}{d \nu} = \frac{N_{\rm ann}^{\rm flare}}{4 \pi d^2} \int dE \frac{d N_e}{dE} \frac{d P}{d \nu}(E),
\eeq
where the power radiated by a single electron is, up to angular factors of order unity, \cite{Rybicki_book}
\beq
\frac{d P}{d \nu}(E) \approx \frac{e^3 B}{m_e c^2}~ F\left(\frac{\nu}{\nu_{\rm s}(E)}\right),
\eeq
where $F(x)$ is a dimensionless function of order unity peaking at $x \approx 0.3$ and $\nu_s(E)$ is the characteristic frequency of radiation given by 
\beq
\nu_{\rm s} \approx \frac{E^2}{(m_e c^2)^3} \frac{e c B}{2 \pi} \approx 1 \textrm{ GHz} ~\left(\frac{E}{10 ~\rm MeV}\right)^2 \frac{B}{1 ~ \rm G}.
\eeq
For simplicity we assume a flat electron--positron spectrum: $E(dN_e/dE) = 2$ for $E \leq m_{\chi} c^2$ (this normalization ensures that the total energy is $2 m_\chi c^2$). We then arrive at
\beq
\frac{d F}{d \nu} = \frac{N_{\rm ann}^{\rm flare}}{4 \pi d^2} \frac{e^3 B}{m_e c^2} G\left(\frac{\nu}{\nu_{\rm s}(m_\chi c^2)}\right),
\eeq
where $G(x) \equiv \int_x^\infty dy/y ~ F(y)$. The function $G$ is of order unity and decays exponentially with characteristic decay scale $x \sim 1$. Numerically, we get
\beq
\frac{d F}{d \nu} \approx 10 ~\textrm{mJy} ~(d_{100})^{-2} \frac{N_{\rm ann}^{\rm flare}}{10^{51}} \frac{B}{1~\textrm{G}}
\eeq
for frequencies $\nu \lesssim \nu_s(m_\chi c^2)$. This flux level is within the reach of existing transient radio surveys (see e.g. Ref.~\cite{Thyagarajan_11}). It would therefore be interesting to make a more quantitative prediction for the time evolution of the synchrotron spectrum for arbitrary ratios $\Delta t_{\rm tce}/t_{\rm syn}$ and more realistic models for the electron--positron spectrum. We defer studying these issues to future work.

\subsection{Gravitational wave bursts}

A natural consequence of TCEs of DMCs is the emission of gravitational waves (GWs), which will be radiated regardless of the microphysical properties of the dark matter. 

The first source of GWs is the orbital motion itself \cite{Kobayashi_04}. For large $\beta$ a burst of GWs is emitted near pericenter passage, with characteristic timescale
\beq
t_{\rm orb} \sim \left(\frac{G M_{\rm BH}}{R_p^3}\right)^{-1/2} \sim T_{\rm cl}\times \beta^{-3/2}.\label{eq:torb}
\eeq
The GW strain due to orbital motion is 
\beq
h_{\rm orb} \sim \frac{G M_{\rm cl} R_p^2}{c^4 d~ t_{\rm orb}^2} \sim \frac{G M_{\rm cl} R_p^2}{c^4 d~ T_{\rm cl}^2} \beta^3, 
\eeq
where $d$ is the distance to the source. 

The tidal compression of the DMC leads to an additional burst of GWs, on a timescale $\Delta t_{\rm tce}$. As shown in Ref.~\cite{Stone_2013}, GW emission can be significantly enhanced for an object that is not perfectly symmetric about the orbital plane. Whereas the asymmetry parameter is expected to be small for stars considered in Ref.~\cite{Stone_2013}, we expect DMCs to be generically triaxial objects \cite{Bardeen_86}. Since our detailed treatment in Section \ref{sec:theory} assumed a perfectly spherical DMC, we shall only use the characteristic lengthscales and timescales derived there to estimate the GW emission during a TCE.

We saw in Section \ref{sec:theory} that the clump is not (yet) significantly deformed in the orbital plane near pericenter passage. We shall therefore assume that $x \sim y \sim R_{\rm cl}$. On the other hand, the clump is significantly compressed perpendicular to the orbital plane, with $z \sim R_{\rm cl}/\beta$ at maximal compression. The components of the reduced quadrupole moment tensor  are therefore of order
\bea
Q_{xx} \sim Q_{xy} \sim Q_{yy} &\sim&  M_{\rm cl} R_{\rm cl}^2, \\
Q_{xz} \sim Q_{yz} &\sim&  M_{\rm cl} R_{\rm cl}^2 /\beta,\\
Q_{zz} &\sim& M_{\rm cl} R_{\rm cl}^2/\beta^2.
\eea
The timescale for compression in the $z$-direction is $\Delta t_{\rm tce} \sim T_{\rm cl} /\beta^2$, while the evolution in the plane is much slower (see Fig.~\ref{fig:comp_XY}). The second derivative of the reduced quadrupole moment is therefore of order
\bea
\ddot{Q}_{xz} \sim \ddot{Q}_{y z} &\sim& \frac{M_{\rm cl} R_{\rm cl}^2}{T_{\rm cl}^2} ~\beta^{3},\\
\ddot{Q}_{zz} &\sim& \frac{M_{\rm cl} R_{\rm cl}^2}{T_{\rm cl}^2} ~\beta^{2},
\eea 
while the other components are subdominant for large $\beta$. Note that the scalings with $\beta$ are different than those derived in Ref.~\cite{Stone_2013} since in their study the star's internal pressure causes a bounce at maximum compression, whereas in our case it is the initial random velocities of the virialized clump that prevent it from collapsing to a point. 

The observed gravitational wave amplitude depends of course on the orientation of the detector with respect to the orbital plane. For generic orientations and large $\beta$, the terms $\ddot{Q}_{xz}$ and $\ddot{Q}_{yz}$ dominate. We therefore get a characteristic GW strain
\bea
h_{\rm tce} \sim \frac{G \ddot{Q}}{c^4 d} \sim  \frac{G M_{\rm cl} R_{\rm cl}^2}{c^4 ~d ~T_{\rm cl}^2} ~\beta^3
\eea
This is a factor $\sim (R_{\rm cl}/R_p)^2$ smaller than the strain due to the orbital motion itself. However, the timescale $\Delta t_{\rm tce}$ for the GW flare from the TCE is shorter by a factor $\sim 1/\sqrt{\beta}$ than the time of pericenter passage \eqref{eq:torb}. The resulting GWs are therefore emitted at frequencies $\sim \sqrt{\beta}$ higher. Numerically, the characteristic frequency of the GW emission due to the TCE is 
\beq
\nu_{\rm tce} \sim 2 \pi \Delta t_{\rm tce}^{-1} \sim 30~\textrm{Hz} \left(\frac{\rho_{\rm cl}}{1~ \textrm{g cm}^{-3}}\right)^{1/2} (\beta/100)^2, \label{eq:nuGW}
\eeq
and the amplitude of the strain is of order
\bea
h_{\rm tce} \sim 10^{-23} ~ \left(\frac{M_{\rm cl}}{0.1~ M_{\odot}} \right)^{5/3} \left(\frac{\rho_{\rm cl}}{1~ \textrm{g cm}^{-3}}\right)^{1/3} \frac{(\beta/100)^3}{d_{100}}.~~~\label{eq:hGW}
\eea
Consider for instance a 0.1 $M_{\odot}$ DMC with density 1 g cm$^{-3}$, tidally compressed by a $10^5 M_{\odot}$ SMBH. The tidal radius is approximately 50 solar radii ($3 \times 10^7$ km), and a penetration factor $\beta = 100$ corresponds to pericenter $R_p \approx 0.5 R_{\odot} \approx R_{\rm cl} \approx R_{\rm BH}$. In this case the amplitude of the GW burst generated by the TCE is as large as that due to the orbital motion, though of course our calculation breaks down both close to the SMBH horizon and when the SMBH--DMC separation becomes comparable to the size of the DMC. From Eqs.~\eqref{eq:nuGW} and \eqref{eq:hGW} we see that with these parameters the burst of GWs from the TCE could be detected by Advanced LIGO and Advanced Virgo \cite{Adligo15}, while that from the orbital motion would remain beyond their reach due to its lower frequency. A detailed computation of the total waveform and its detectability by current or future GW detectors is beyond the scope of this work. We therefore simply point out that TCEs of DMCs ought to produce a unique GW signal, that could make it possible to distinguish them from signatures of other close encounters with SMBHs.

\section{Discussion}
\label{sec:discussion}

In this paper we have outlined the essential physics of TCEs of DMCs and predicted several observational consequences. Our calculations rely on a rather simplified model, however, and we list below several points that need to be explored in more detail to obtain more precise quantitative estimates.

{\it Different clump profiles.}
In our calculation of the DMC compression, we have assumed a simple phase-space density, with a Gaussian density profile. Our results can in principle be generalized to more realistic density profiles, such as the isothermal profile $\rho \propto r^{-2}$ or the NFW profile \cite{NFW_1997}. Since most profiles do not generally factor into independent functions along each axis, the calculation would not simplify as it did for a Gaussian profile, but the formalism remains identical. A more difficult aspect would be to consistently follow different layers of the DMC as their tidal radii are different. This is particularly important for a cusped profile. In this case one ought to numerically solve for the trajectories of the clump particles under the combined gravitational pull of the DMC at that of the SMBH. Simulations may provide a useful tool to estimate the effect of different clump profiles.

{\it Self-gravity.}
We have assumed that the DMC self gravity can be entirely neglected immediately at the crossing of the tidal radius, and that the later compression of the DMC does not affect this result. In reality, of course, the self-gravity of the DMC ought to be self-consistently included, and may significantly affect our results. Near pericenter passage where the DMC is highly elongated and compressed, instabilities may develop due to self gravity and the clump could fragment. 

{\it Self-interactions.}
We have only considered collisionless DM in this work. The dynamical behavior of the DMC under compression would be modified if the DM is self-interacting \cite{Kaplinghat_15}.

{\it Tidal approximation.}
We have restricted our calculation to the tidal approximation, for which the DMC extent is much smaller than its separation from the SMBH. It would be interesting to generalize this calculation to close encounters / extended DMCs.

{\it Background star cluster.}
We have assumed that the SMBH dominates the gravitational potential, neglecting the effect of the background star cluster. This could have an effect if the tidal radius is comparable to that of the SMBH sphere of influence. As long as the stellar cluster is spherically symmetric, the form of the equations derived in this work should be unchanged: the center-of-mass orbit will still be planar, and the in- and out-of-plane deviations should still be separable and linearizable provided $R_{cl}$ is much smaller than the characteristic lengthscale of the potential.

{\it Newtonian approximation.}
Throughout this calculation we have taken a Newtonian approximation, ignoring general-relativistic effects. If the DMC approaches the SMBH within a few Schwarzschild radii, a fully relativistic treatment is required. Interesting effects may arise if the SMBH is rapidly spinning \cite{Kesden_12}.

{\it Rate estimates.} The obvious next step of this work is to estimate the rates of TCEs of DMCs. This requires $(i)$ a prescription for the abundance and mass function of DMCs, accounting for their tidal destruction by stars and galactic tides and $(ii)$ a more detailed study of the loss-cone problem of DMCs and the distribution of penetration factors $\beta$. The rates may be significantly enhanced in binary SMBHs \cite{Wegg_11}. We leave these important questions to future work.

\section{Conclusion}
\label{sec:conclusion}

We have calculated the tidal deformation and compression of a DMC penetrating into the tidal sphere of a SMBH. We found that this process results in extreme compression in the direction perpendicular to the orbital plane, with a duration and compression factor that depend on the initial density of the DMC core, as well as on the ratio between the pericenter radius of the DMC orbit and the tidal radius of the SMBH. As a result of the boost in density and velocity dispersion, a natural signature of this model is in the form of flares (of gamma-rays, for example) from annihilating dark-matter particles in the clump. The amplitude of these annihilation flares relative to the quiescent DMC annihilation rate is particularly pronounced for $p$- and $d$-wave annihilation.

Under the WIMP scenario, we calculated the characteristic amplitude and duration of gamma-ray flares, covering the full parameter space of DMC mass and core density, dark-matter particle mass, and orbital penetration factors. 
Comparing our predictions with observations from Fermi LAT \cite{Fermi_2013}, we found that flares recently detected in the data, some with no known counterparts in the point-source catalogs, are consistent with those expected from TCEs of DMCs for a range of model parameters.   
We emphasize that our model further predicts that the flares should be distributed isotropically on the sky, and exhibit a universal energy spectrum. Our results therefore motivate a more exhaustive search for flares in Fermi LAT data, including ones on shorter timescales, the finding of which may enable a more detailed comparison against the model presented here, as well as many alternative scenarios. 

We also addressed other possible signatures of TCEs of DMCs.  We derived the characteristic flux of synchrotron radiation that would be produced if the products of dark matter annihilation are relativistic positrons and electrons. We also discussed the gravitational-wave signature resulting from a TCE and argued that it is detectable by Advanced LIGO for certain parameters.

We mentioned several caveats and discussed possible generalizations and improvements to this model. Looking forward, dedicated numerical simulations would be a particularly interesting and useful follow-up.

While we have focused on the characteristics of TCEs of DMCs given arbitrary clump parameters, more work lies ahead before we can use these results to constrain the properties of dark matter. The next steps are, first, to derive the clump parameters and their distribution from any particular model for the small-scale DM power spectrum, and second, to make predictions for the event rate, accounting for stellar encounters and loss-cone physics. It is our hope that the interesting new observables we have introduced in this paper will stimulate more work in this direction.

\section*{Acknowledgments}
We thank Ilias Cholis, Adrienne Erickcek, Marc Kamionkowski and Tristan Smith for useful discussions. YAH thanks Scott Tremaine, Matias Zaldarriaga, and members of the Institute for Advanced Study for valuable comments. This work was supported by the John Templeton Foundation. JS also acknowledges support by ERC project 267117 (DARK) hosted by UPMC, and  by NSF grant OIA-1124403.

\appendix

\section{Clump deformation in the orbital plane}\label{app:in-plane}

Here we derive Eq.~\eqref{eq:rho_inplane}, following the same steps as in Section \ref{sec:z} but generalizing it to a two-by-two-dimensional phase-space. We use standard linear algebra results for the inverse and determinant of a block matrix.

The components of $\bs{r}$ and $\bs{v}$ in the orbital plane depend linearly on initial conditions:
\bea
\bs{r}_{||}(f) &=& \bs{A}(f) \bs{r}_{||0} + \bs{B}(f)  \bs{v}_{||0}, \label{eq:r_of_r0v0}\\
\bs{v}_{||}(f) &=& \bs{\dot{A}}(f)  \bs{r}_{||0} + \bs{\dot{B}}(f) \bs{v}_{||0}.
\eea

The initial positions and velocities can be obtained with the inverse transform 
\bea
\bs{r}_{||0} &=& \bs{\tilde{A}} \bs{r}_{||} + \bs{\tilde{B}} \bs{v}_{||}, \label{eq:x||0_of_r}\\
\bs{v}_{||0} &=& \bs{\tilde{C}} \bs{r}_{||} + \bs{\tilde{D}} \bs{v}_{||}, \label{eq:v||0_of_r}
\eea
with the following explicit expressions for $\bs{\tilde{B}}, \bs{\tilde{C}}, \bs{\tilde{D}}$, if $\bs{A}$ is non-singular (we will not need an expression for $\bs{\tilde{A}}$):
\bea
\bs{\tilde{D}} &\equiv& (\bs{\dot{B}} - \bs{\dot{A}} \bs{A}^{-1} \bs{B})^{-1}, \label{eq:tildeB}\\
\bs{\tilde{B}} &\equiv& - \bs{A}^{-1} \bs{B} \bs{\tilde{D}} , \label{eq:tildeD}\\
\bs{\tilde{C}} &\equiv& - \bs{\tilde{D}} \bs{\dot{A}}\bs{A}^{-1}.
\eea
The determinant of the $(\bs{r}_{||0}, \bs{v}_{||0}) \rightarrow (\bs{r}_{||}, \bs{v}_{||})$ linear transformation is unity. Explicitly, this determinant is
\be
\det(\bs{A}) \det(\bs{\dot{B}} - \bs{\dot{A}} \bs{A}^{-1} \bs{B}) = 1,
\ee
which implies that $\det \bs{\tilde{D}} = \det{\bs{A}}$. We now rewrite Eq.~\eqref{eq:r_of_r0v0} as
\be
\bs{r}_{||0}= \bs{A}^{-1} \bs{r}_{||} - \bs{A}^{-1} \bs{B} \bs{v}_{||0}.
\ee
Using Eq.~\eqref{eq:v||0_of_r} we obtain the determinant of the $\bs{v}_{||} \rightarrow \bs{v}_{||0}$ change of variables at constant $\bs{r}_{||}$:
\be
d^2 v_{||} = \frac{d^2 v_{||0}}{|\det{\bs{\tilde{D}}}|} = \frac{d^2 v_{||0}}{|\det{\bs{A}}|}.
\ee
The contribution of the in-plane axes to Eq.~\eqref{eq:rho} is therefore
\bea
\tilde{\rho}_{||}(\bs{r}_{||}, f) = \rho_{*} ^{2/3} \frac3{2 \pi} \int \frac{d^2 v_{||0}}{|\det{\bs{A}}|} \rme^{-3 \bs{v}_{||0}^2/2}\\
\rme^{- 3(\bs{A}^{-1} \bs{r}_{||} - \bs{A}^{-1} \bs{B} \bs{v}_{||0})^2/2}.
\eea
After some linear algebra, we compute the Gaussian integral and arrive at Eq.~\eqref{eq:rho_inplane}.

\bibliography{dm_clumps.bib}

\end{document}